\documentclass[10pt,aps,prd,twocolumn,nofootinbib,superscriptaddress]{revtex4-2}
\usepackage{amsmath,amssymb,mathtools,bm}
\usepackage{graphicx}
\usepackage{booktabs}
\usepackage{siunitx}
\usepackage{hyperref}
\usepackage{microtype}
\hypersetup{colorlinks=true,citecolor=blue,linkcolor=blue,urlcolor=blue}
\usepackage{orcidlink}
\newcommand{\ii}{\mathrm{i}}
\newcommand{\TT}{\mathrm{TT}}
\newcommand{\cH}{\mathcal H}
\newcommand{\Om}{\Omega_{\rm m}}
\newcommand{\Ol}{\Omega_{\Lambda}}
\newcommand{\dd}{\mathrm{d}}

\begin{document}

\title{Finite-Momentum Kinetic Corrections to Viscous Tensor Perturbations in an Expanding Universe}
 
\author{Nishil Savla, \orcidlink{0009-0002-0881-9814}}
\email{savla.nishil03@gmail.com}
\affiliation{Department of Physics, St. Xavier's College (Autonomous), Ahmedabad, Gujarat, India}

\author{Gurudatt Gaur}
\email{gurudattgaur@gmail.com}
\affiliation{Department of Physics, St. Xavier's College (Autonomous), Ahmedabad, Gujarat, India}

\date{\today}

\begin{abstract}
We study tensor perturbations propagating through a viscous relativistic medium in a spatially flat FLRW universe, with particular emphasis on the correction produced by spatial free streaming beyond a local causal relaxation model. We start from the relaxation-time Boltzmann equation used by Baym, Patil, and Pethick to describe the response of matter to a gravitational wave. In the zero-streaming limit, the tensor stress obeys a Maxwell--Cattaneo (or linear Müller--Israel--Stewart, MIS-type) relation. For the single-relaxation-time model considered here, the corresponding macroscopic relaxation time is identified with the microscopic relaxation time, $\tau_\pi=\tau_c$.

We then retain the spatial streaming term. For an ultrarelativistic isotropic medium, the angular response can be evaluated analytically,
\begin{equation}
F(z)=\frac{15}{16}\int_{-1}^{1}
\frac{(1-\mu^2)^2}{1-z\mu}\,\dd\mu
=\sum_{n=0}^{\infty}
\frac{15z^{2n}}{(2n+1)(2n+3)(2n+5)}.
\end{equation}
The resulting finite-momentum response is
\begin{equation}
R_{\rm kin}(x)
=\frac{F\!\left(x/(x+\ii)\right)}{1-\ii x},
\qquad x=\omega\tau_c,
\end{equation}
for local propagation with $q=\omega$ and particle speed $v=1$. The local MIS response is recovered by setting $F=1$.

We solve the tensor Einstein equation together with a discretized relaxation-time Boltzmann equation and compare the result with the local MIS propagation model. The calculation is performed in a flat matter-plus-$\Lambda$ background. The analytic angular response agrees with direct angular quadrature at relative errors below $4\times10^{-12}$. For the illustrative normalization used here, the full numerical calculation gives a nonmonotonic correction to the tensor power transfer function, with a maximum of about $0.8\%$ near $x\simeq0.37$ and a minimum of about $-5.6\%$ near $x\simeq1.97$. The locations and sizes are stable under angular-resolution and ODE-tolerance tests and are reproduced by the kinetic WKB calculation.

The finite-momentum correction is a property of the specified relaxation-time kinetic model and is not claimed to be universal. In particular, the high-frequency local-MIS asymptote cannot be extrapolated into the regime where the full kinetic response, including streaming and in an expanding background possible Landau damping, is required.
\end{abstract}

\maketitle

\section{Introduction}

Gravitational waves can travel through matter before reaching an observer. If the matter has a shear response, the wave can transfer energy to the medium, changing both its amplitude and phase. The effect becomes particularly relevant when the wave period is comparable to a microscopic relaxation time.

A simple description uses ordinary viscous hydrodynamics. In this limit the transverse-traceless anisotropic stress is proportional to the shear rate,
\begin{equation}
\pi_{ij}^{\TT}=-\eta\dot h_{ij},
\end{equation}
where $\eta$ is the shear viscosity. With the conventions used below, this gives the familiar hydrodynamic damping rate
\begin{equation}
\gamma_{\rm hyd}=16\pi G\eta.
\end{equation}
This result follows from kinetic theory in the analysis of gravitational-wave damping by Baym, Patil, and Pethick \cite{Baym2017}, and the corresponding hydrodynamic attenuation was used by Goswami and collaborators to constrain cosmological viscosity with gravitational-wave observations \cite{Goswami2016,Goswami2017}.

Ordinary viscous hydrodynamics assumes that the matter response is local in time. A microscopic medium has a finite relaxation time, so the stress has memory. Israel and Stewart introduced causal relativistic theories in which dissipative variables have their own relaxation equations \cite{Israel1979}. In the tensor sector, the local linearized relation considered here is
\begin{equation}
\tau_\pi\dot\pi_{ij}^{\TT}+\pi_{ij}^{\TT}
=-\eta\dot h_{ij}.
\label{eq:MIS}
\end{equation}
The relation is used here as a local subhorizon constitutive law, not as a complete covariant MIS theory. Recent work has emphasized both the microscopic accuracy and the restricted domain of transient Israel--Stewart descriptions \cite{Wagner2024,Gavassino2024}. The response of relativistic fluids to metric perturbations has also been derived directly in linear response within MIS hydrodynamics \cite{Stoetzel2025}.

The present work has a narrower goal. We start from the relaxation-time Boltzmann response and ask what changes when the spatial streaming term is retained. The local MIS relation is obtained by neglecting spatial free streaming. The exact expansion parameter is
\begin{equation}
\epsilon_{\rm fs}
=
\frac{qv\tau_c}
{\sqrt{1+\omega^2\tau_c^2}},
\label{eq:epsfs}
\end{equation}
where $q$ is the local wave number, $v$ is the particle speed, and $\tau_c$ is the microscopic collision time. The commonly used condition $qv\tau_c\ll1$ is a sufficient, stronger condition for this expansion.

The main calculation has four parts. First, we write the tensor Einstein equation in an FLRW background. Second, we derive the local relaxation response and identify its MIS limit. Third, we evaluate the finite-momentum angular response analytically. Finally, we solve the coupled Boltzmann and Einstein equations numerically and compare the result with the corresponding WKB propagation.

The numerical calculation is used to distinguish a physical finite-momentum effect from numerical artifacts. We therefore perform independent angular-integration tests, angular-resolution tests, ODE-tolerance tests, and subhorizon convergence tests.

\section{Tensor perturbations in FLRW}

We consider a spatially flat FLRW metric with a transverse-traceless perturbation,
\begin{equation}
\dd s^2=a^2(\eta)
\left[-\dd\eta^2+(\delta_{ij}+h_{ij})\dd x^i\dd x^j\right],
\end{equation}
where
\begin{equation}
\partial_i h_{ij}=0,
\qquad
h_{ii}=0.
\end{equation}
A prime denotes differentiation with respect to conformal time $\eta$, and
\begin{equation}
\cH\equiv\frac{a'}{a}
\end{equation}
is the conformal Hubble parameter. The physical Hubble parameter will be denoted by $H=\dot a/a$. The conformal and physical Hubble parameters are related by $\cH=aH$.

At linear order the tensor Einstein equation can be written as
\begin{equation}
h_{ij}''+2\cH h_{ij}'+q^2h_{ij}
=
16\pi G a^2\pi_{ij}^{\TT}.
\label{eq:Einstein}
\end{equation}
Here $q$ is the comoving wave number, while the physical wave number is $k_{\rm phys}=q/a$. In a locally normalized Minkowski patch with $a=1$, these coincide. The sign convention for $\pi_{ij}^{\TT}$ is chosen so that the Navier--Stokes relation gives positive damping.

For a mode well inside the cosmological horizon,
\begin{equation}
q\gg\cH,
\end{equation}
the vacuum solution has the leading WKB form
\begin{equation}
h_{ij}\propto a^{-1}e^{\ii q\eta}.
\end{equation}
The factor $1/a$ is the usual dilution of a freely propagating tensor mode. Matter produces an additional slowly varying amplitude and phase.

\section{Relaxation-time Boltzmann response}

The microscopic starting point is the relaxation-time Boltzmann equation. Baym et al.\ used this framework to calculate the response of matter to a weak gravitational wave \cite{Baym2017}. In a local Minkowski region, the perturbation of the particle energy produced by a tensor perturbation is
\begin{equation}
\delta\epsilon
=
-\frac{1}{2}h_{ij}\frac{p_i p_j}{\epsilon_0}.
\end{equation}

For a Fourier mode with convention $e^{\ii(\mathbf q\cdot\mathbf r-\omega t)}$, the linearized relaxation-time equation contains the denominator
\begin{equation}
\omega-\mathbf q\cdot\mathbf v+\frac{\ii}{\tau_c}.
\label{eq:denominator}
\end{equation}
The term $\mathbf q\cdot\mathbf v$ is the spatial streaming term. If it is neglected, the response is local in space. If it is retained, particles move through the wave during the relaxation process.

The tensor stress can be written schematically as
\begin{equation}
\pi_{ij}(\mathbf q,\omega)
=
\int_{\mathbf p}
{\cal W}_{ij}(\mathbf p)
\frac{\omega}
{\mathbf q\cdot\mathbf v-\omega-\ii/\tau_c}
\frac{\partial f_0}{\partial\epsilon_0},
\label{eq:stressschematic}
\end{equation}
where ${\cal W}_{ij}$ contains the tensor momentum weight and the TT projection is understood. The collision-dominated limit is
\begin{equation}
\omega\tau_c\ll1,
\qquad
qv\tau_c\ll1.
\end{equation}
In this limit the usual viscous relation is recovered \cite{Baym2017}.

For a general collision operator the relaxation time appearing in a macroscopic tensor constitutive equation need not be identical to a microscopic collision time. In the single-relaxation-time model used here, however, the local tensor response obtained by setting $q=0$ has the same relaxation denominator as Eq.~\eqref{eq:denominator}. We therefore identify
\begin{equation}
\boxed{\tau_\pi=\tau_c\equiv\tau}
\label{eq:tauidentification}
\end{equation}
throughout this calculation. This identification is model-specific; it is not asserted for arbitrary microscopic collision kernels.

\section{Local MIS response}

Setting the spatial streaming term to zero gives the local causal response
\begin{equation}
\tau\dot\pi_{ij}^{\TT}+\pi_{ij}^{\TT}
=
-\eta\dot h_{ij}.
\label{eq:localMIS}
\end{equation}
For a harmonic perturbation with time dependence $e^{-\ii\omega t}$,
\begin{equation}
\pi_{ij}^{\TT}
=
-\frac{\eta}{1-\ii\omega\tau}\dot h_{ij}.
\end{equation}
Define
\begin{equation}
x=\omega_{\rm phys}\tau.
\label{eq:xdef}
\end{equation}
The dimensionless local response factor is
\begin{equation}
R_{\rm MIS}(x)
=
\frac{1}{1-\ii x}
=
\frac{1}{1+x^2}
+\ii\frac{x}{1+x^2}.
\label{eq:Rmis}
\end{equation}
The Navier--Stokes limit is recovered for $x\ll1$. At high frequency the magnitude of the local response decreases as $x^{-1}$.

The local dispersion relation follows by inserting the constitutive response into the subhorizon tensor equation. For a locally monochromatic mode,
\begin{equation}
h\propto e^{-\ii\omega t},
\end{equation}
the matter-induced frequency shift, to first order in the small fractional shift $|\delta\omega|/\omega\ll1$, is
\begin{equation}
\delta\omega
=
8\pi G\eta\,
\frac{x-\ii}{1+x^2}.
\label{eq:domega}
\end{equation}
Thus
\begin{equation}
\delta\omega_R
=
8\pi G\eta\frac{x}{1+x^2},
\qquad
\delta\omega_I
=
-8\pi G\eta\frac{1}{1+x^2}.
\end{equation}
The negative imaginary part gives attenuation. If $\gamma_{\rm MIS}$ denotes the energy-like damping rate, then
\begin{equation}
\boxed{
\gamma_{\rm MIS}
=
\frac{16\pi G\eta}{1+x^2}.
}
\label{eq:gammaMIS}
\end{equation}
The corresponding phase-rate magnitude is
\begin{equation}
\boxed{
\delta\omega_R
=
8\pi G\eta\frac{x}{1+x^2}.
}
\label{eq:phaseRate}
\end{equation}
The factor of two between the energy-like damping rate and the amplitude attenuation exponent is conventional.

\section{Finite-momentum kinetic correction}

The local MIS approximation removes the streaming term from Eq.~\eqref{eq:denominator}. To retain it, define
\begin{equation}
A=\omega+\frac{\ii}{\tau_c}.
\end{equation}
Writing $\mu=\cos\theta$ for the angle between the particle velocity and the wave vector, the angular denominator is $A-qv\mu$. For small streaming,
\begin{equation}
\frac{1}{A-qv\mu}
=
\frac{1}{A}
+\frac{qv\mu}{A^2}
+\frac{q^2v^2\mu^2}{A^3}
+\cdots.
\end{equation}
For an isotropic equilibrium distribution the odd term vanishes after angular integration. Hence the first nonzero correction is quadratic in the streaming parameter,
\begin{equation}
\frac{\Delta\pi_{\rm kin}}{\pi_{\rm MIS}}
=
\mathcal O(\epsilon_{\rm fs}^2),
\qquad
\epsilon_{\rm fs}
=
\frac{qv\tau_c}
{\sqrt{1+\omega^2\tau_c^2}}.
\label{eq:streamingexpansion}
\end{equation}
The stronger condition $qv\tau_c\ll1$ is sufficient but is not the most general expansion criterion.

For the numerical model we take an ultrarelativistic isotropic medium with $v=1$ and use the local gravitational-wave dispersion relation $q=\omega$. The angular response can then be normalized independently of the radial momentum integral. The relevant tensor angular function is
\begin{equation}
F(z)
=
\frac{15}{16}
\int_{-1}^{1}
\frac{(1-\mu^2)^2}{1-z\mu}\,\dd\mu.
\label{eq:F}
\end{equation}
The normalization satisfies
\begin{equation}
\frac{15}{16}
\int_{-1}^{1}(1-\mu^2)^2\,\dd\mu
=1.
\end{equation}

Expanding the denominator gives
\begin{equation}
\boxed{
F(z)
=
\sum_{n=0}^{\infty}
\frac{15z^{2n}}
{(2n+1)(2n+3)(2n+5)}.
}
\label{eq:Fseries}
\end{equation}
The first terms are
\begin{equation}
F(z)
=
1+\frac{z^2}{7}
+\frac{z^4}{21}
+\frac{5z^6}{231}
+\frac{5z^8}{429}
+\mathcal O(z^{10}).
\end{equation}
The exact expression is
\begin{equation}
F(z)
=
\frac{5}{16z^5}
\left[
10z^3-6z
+3(z^2-1)^2
\ln\left(\frac{1+z}{1-z}\right)
\right],
\label{eq:Fexact}
\end{equation}
with the logarithm understood on the analytic branch reached by continuation from $z=0$. The values of $z$ used below do not cross the logarithmic branch cut.

For $v=1$ and $q=\omega$, define
\begin{equation}
x=\omega\tau_c.
\end{equation}
Then
\begin{equation}
z
=
\frac{qv}{\omega+\ii/\tau_c}
\tau_c
=
\frac{x}{x+\ii},
\end{equation}
and the exact angularly averaged response relative to the Newtonian tensor normalization is
\begin{equation}
\boxed{
R_{\rm kin}(x)
=
\frac{F\!\left(x/(x+\ii)\right)}
{1-\ii x}.
}
\label{eq:Rkin}
\end{equation}
The local MIS result is recovered when the streaming factor is replaced by $F=1$.

Since
\begin{equation}
\frac{R_{\rm kin}}{R_{\rm MIS}}
=
F\!\left(\frac{x}{x+\ii}\right),
\end{equation}
the low-frequency expansion gives
\begin{equation}
\operatorname{Re}
\left(
\frac{R_{\rm kin}}{R_{\rm MIS}}-1
\right)
=
-\frac{x^2}{7}
+\mathcal O(x^4).
\label{eq:lowkin}
\end{equation}
The numerical calculation verifies this coefficient: at $x=10^{-3}$ the ratio of the numerical correction to $-x^2/7$ is $0.999994$.

At high frequency,
\begin{equation}
\frac{x}{x+\ii}\longrightarrow1,
\end{equation}
and
\begin{equation}
F(1)=\frac54.
\end{equation}
Therefore
\begin{equation}
\left|
\frac{R_{\rm kin}}{R_{\rm MIS}}
\right|
\longrightarrow\frac54.
\end{equation}
This is a property of the specified angular RTA model. It should not be interpreted as the full collisionless response, because the latter requires the complete kinetic treatment and, in an expanding background, can include Landau damping \cite{Baym2017}.

\section{Propagation in a flat matter-plus-Lambda background}

We use a spatially flat matter-plus-$\Lambda$ background,
\begin{equation}
H(a)
=
H_0\sqrt{\Om a^{-3}+\Ol},
\qquad
\Om+\Ol=1,
\label{eq:H}
\end{equation}
where $H=\dot a/a$ is the physical Hubble parameter.

The physical wave frequency is
\begin{equation}
\omega_{\rm phys}=\frac{q}{a}.
\end{equation}
For the conformal scaling
\begin{equation}
\tau(a)=\tau_0 a,
\label{eq:tauscaling}
\end{equation}
the relaxation parameter is constant along the ray:
\begin{equation}
x
=
\omega_{\rm phys}\tau(a)
=
q\tau_0.
\label{eq:xconstant}
\end{equation}
The scaling in Eq.~\eqref{eq:tauscaling} is an assumption corresponding to a relaxation time proportional to the inverse temperature in an approximately conformal relativistic medium. It is not a universal property of arbitrary cosmological fluids.

For comparison with the frequency-independent hydrodynamic result, define
\begin{equation}
D_{\rm G}
=
8\pi G
\int_{a_s}^{1}
\frac{\eta(a)}{aH(a)}\,\dd a.
\label{eq:DG}
\end{equation}
The local MIS amplitude optical depth and phase magnitude are then
\begin{equation}
D_{\rm MIS}
=
\frac{D_{\rm G}}{1+x^2},
\qquad
\Delta\Psi_{\rm MIS}
=
D_{\rm G}\frac{x}{1+x^2}.
\label{eq:DMIS}
\end{equation}
With the Fourier convention $e^{-\ii\omega t}$ used in the local dispersion calculation, the transfer factor is
\begin{equation}
\boxed{
{\cal T}_{\rm MIS}
=
\exp\left[-\frac{D_{\rm G}}{1+x^2}\right]
\exp\left[-\ii D_{\rm G}\frac{x}{1+x^2}\right].
}
\label{eq:TMIS}
\end{equation}
The standard vacuum factor $1/a$ is separate from this matter-induced transfer factor.

The causal phase and attenuation obey the consistency relation
\begin{equation}
\boxed{
\Delta\Psi_{\rm MIS}
=
xD_{\rm MIS}.
}
\label{eq:phaseconsistency}
\end{equation}
At $x\ll1$,
\begin{equation}
D_{\rm MIS}
=
D_{\rm G}\left(1-x^2+\mathcal O(x^4)\right),
\end{equation}
while
\begin{equation}
\Delta\Psi_{\rm MIS}
=
D_{\rm G}\left(x-x^3+\mathcal O(x^5)\right).
\end{equation}
At $x\gg1$,
\begin{equation}
D_{\rm MIS}\simeq D_{\rm G}x^{-2},
\qquad
\Delta\Psi_{\rm MIS}\simeq D_{\rm G}x^{-1}.
\end{equation}
The phase magnitude is maximal at $x=1$, where
\begin{equation}
D_{\rm MIS}=\frac{D_{\rm G}}{2},
\qquad
\Delta\Psi_{\rm MIS}=\frac{D_{\rm G}}{2}.
\end{equation}

If one additionally assumes the conformal viscosity scaling
\begin{equation}
\eta(a)=\eta_0a^{-3},
\label{eq:etascaling}
\end{equation}
then
\begin{equation}
D_{\rm G}
=
\frac{8\pi G\eta_0}{H_0}
I(z_s),
\end{equation}
where
\begin{equation}
I(z_s)
=
\int_{1/(1+z_s)}^1
\frac{a^{-4}}
{\sqrt{\Om a^{-3}+\Ol}}
\,\dd a.
\end{equation}
The scaling in Eq.~\eqref{eq:etascaling} is likewise a model assumption. It can arise for an approximately conformal relativistic medium with $\eta\propto T^3$, but it is not a generic cosmological law.

\section{Numerical method}

We solve the tensor Einstein equation together with a discretized relaxation-time Boltzmann equation. The angular variable is evaluated with Gauss--Legendre quadrature. The final calculation uses $N_\mu=64$ and tests $N_\mu=16,32,48,80$.

The background parameters are
\begin{equation}
\Om=0.3,
\qquad
\Ol=0.7,
\qquad
z_s=2,
\end{equation}
so that $a_s=1/3$. We choose
\begin{equation}
D_{\rm G}=0.5,
\qquad
k=1000,
\qquad
ka_s=333.3.
\end{equation}
These are numerical normalization choices and are not measurements of the real cosmological viscosity.

The angular normalization is
\begin{equation}
\frac{15}{16}
\int_{-1}^{1}(1-\mu^2)^2\,\dd\mu
=1.
\end{equation}
The analytic angular function agrees with direct quadrature with maximum relative error $3.19\times10^{-12}$.

The main numerical quantity is the power ratio
\begin{equation}
{\cal P}(x)
=
\frac{|T_{\rm kin}(x)|^2}
{|T_{\rm MIS}(x)|^2}.
\label{eq:power}
\end{equation}
A dense scan of 142 values between $x=0.03$ and $x=10$ was used.

The numerical calculation is performed in the single-relaxation-time ultrarelativistic model defined above, so that the same relaxation time $\tau=\tau_c=\tau_\pi$ enters the kinetic and local-MIS responses. This equality is part of the model specification and should not be generalized to arbitrary collision operators without a separate microscopic derivation.

\section{Numerical results}

The full numerical calculation gives a nonmonotonic finite-momentum correction. At low $x$ the correction is small. It then becomes positive and reaches a maximum at
\begin{equation}
x_{\rm max}\simeq0.3733,
\end{equation}
with
\begin{equation}
\boxed{
{\cal P}(x_{\rm max})-1
=
7.979\times10^{-3}.
}
\end{equation}
This corresponds to an enhancement of about $0.8\%$.

The correction then decreases and reaches a minimum near
\begin{equation}
x_{\rm min}\simeq1.97,
\end{equation}
where
\begin{equation}
\boxed{
{\cal P}(x_{\rm min})-1
=
-5.578\times10^{-2}.
}
\end{equation}
The suppression is therefore about $5.6\%$ relative to the local MIS calculation.

The WKB result gives the same locations on the numerical grid. Near the maximum the full correction is $7.979\times10^{-3}$ and the WKB correction is $7.980\times10^{-3}$. Near the minimum the full correction is $-5.578\times10^{-2}$ and the WKB correction is $-5.560\times10^{-2}$.

\begin{figure}[t]
\centering
\includegraphics[width=\columnwidth]{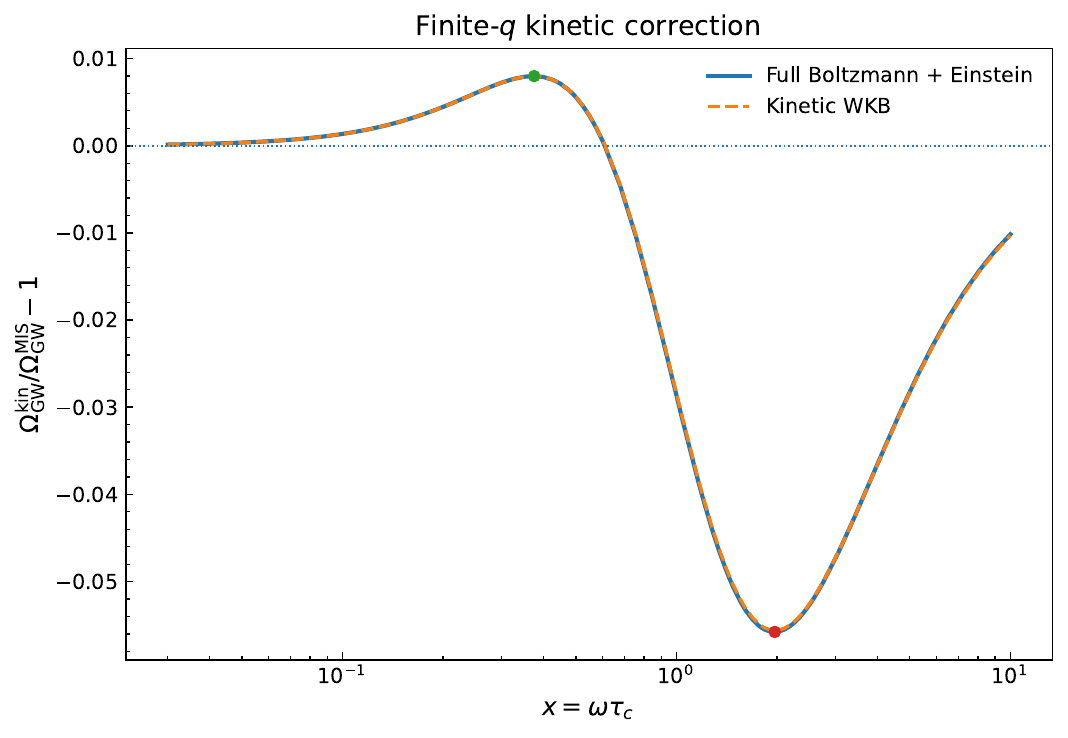}
\caption{Relative kinetic correction to the gravitational-wave power. The full Boltzmann calculation is compared with the kinetic WKB result.}
\label{fig:spectral}
\end{figure}

\begin{figure}[t]
\centering
\includegraphics[width=\columnwidth]{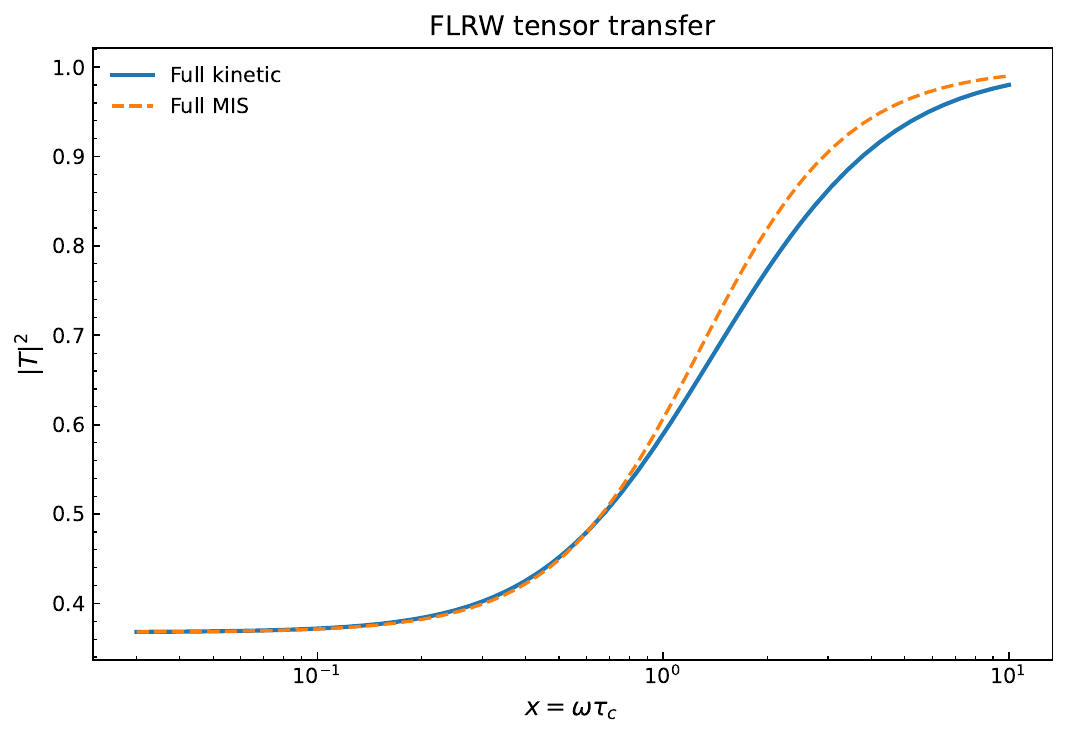}
\caption{Full FLRW tensor transfer functions for the kinetic and MIS calculations.}
\label{fig:transfer}
\end{figure}

\begin{figure}[t]
\centering
\includegraphics[width=\columnwidth]{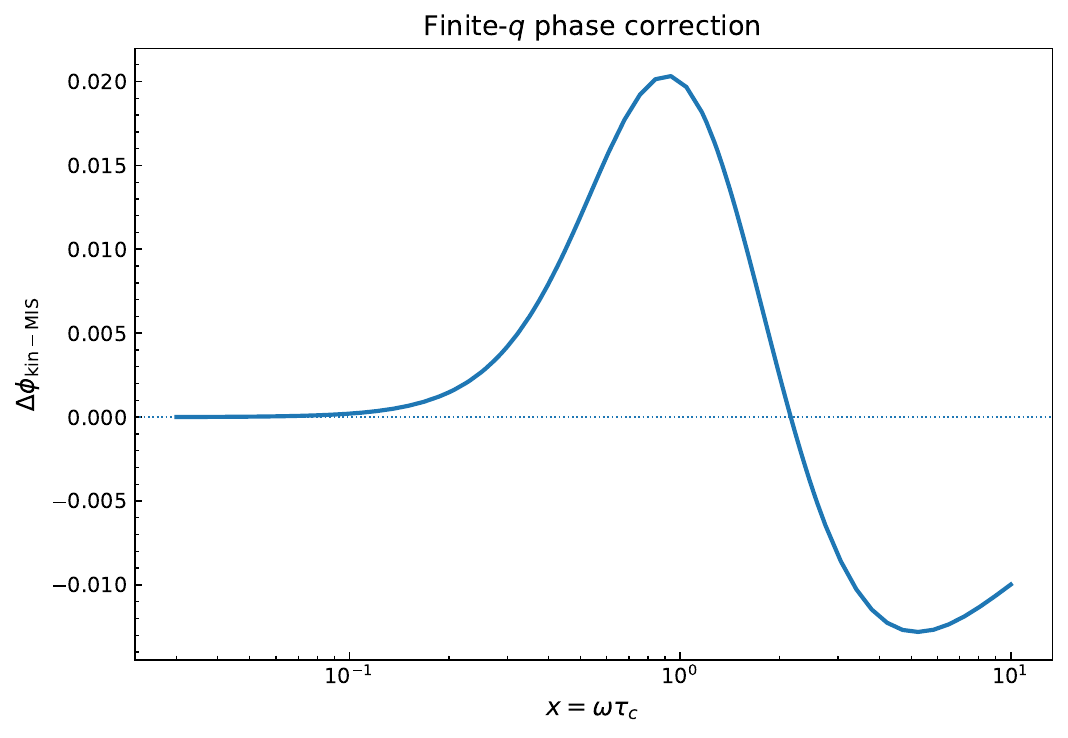}
\caption{Finite-momentum kinetic phase correction relative to the local MIS result.}
\label{fig:phase}
\end{figure}

\section{Numerical validation}

Several tests were performed to check that the correction is not caused by the numerical procedure.

First, the analytic angular response was compared with independent numerical quadrature. The maximum relative error was $3.19\times10^{-12}$.

Second, the angular resolution was changed. At $x=0.376$, the power ratio changed from $1.0079780$ at $N_\mu=16$ to $1.0079779$ at $N_\mu=80$. At $x=1.98$, the result remained near $0.944225$ across the same resolutions.

Third, the ODE tolerance was changed. At $x=0.376$, the power ratio converged from $1.008013$ at a loose tolerance to $1.007978$ at the tightest tolerance used. At $x=1.98$, it converged from $0.944234$ to approximately $0.944225$.

Finally, the full FLRW result was compared with WKB propagation at increasing subhorizon wave number. At $x=0.376$, the amplitude error decreased from $1.12\times10^{-3}$ at $ka_s=66.7$ to $3.01\times10^{-4}$ at $ka_s=266.7$. At $x=1.98$, it decreased from $8.94\times10^{-4}$ to $2.20\times10^{-4}$.

These checks support the use of the WKB expression in the subhorizon regime and show that the nonmonotonic correction is not explained by angular resolution or ODE tolerance.

\begin{figure}[t]
\centering
\includegraphics[width=\columnwidth]{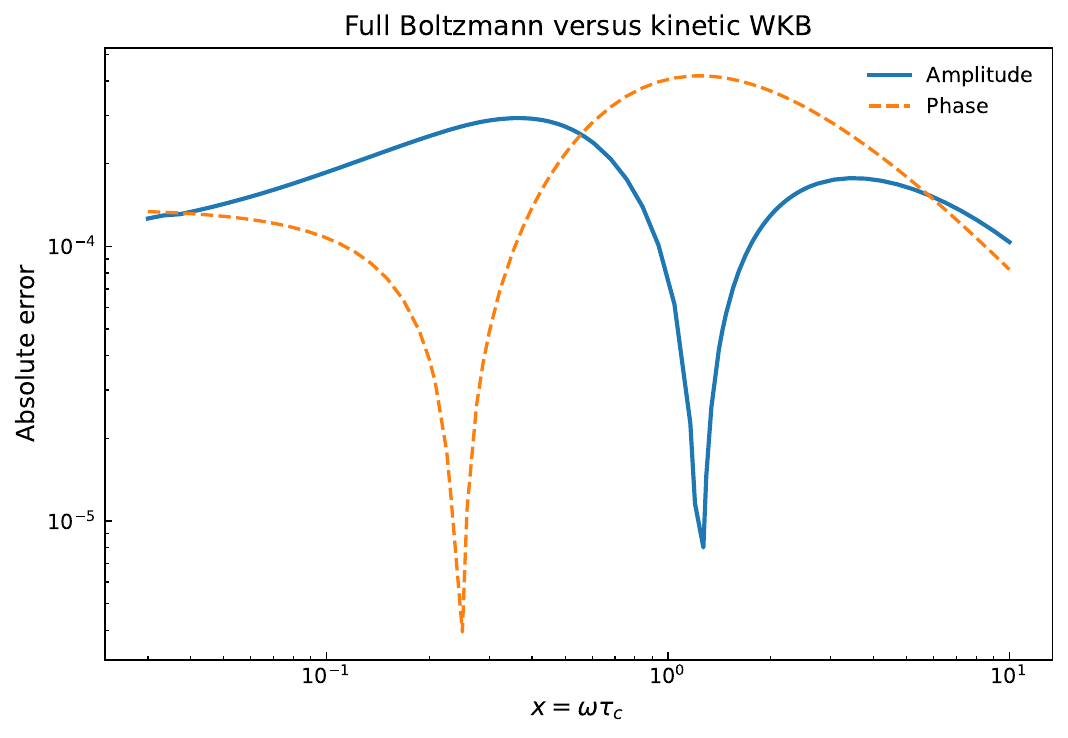}
\caption{Difference between the full Boltzmann evolution and the kinetic WKB result.}
\label{fig:wkb}
\end{figure}

\section{Domain of validity}

The calculation involves several independent approximations, and they should not be conflated.

First, the cosmological propagation is treated in the subhorizon WKB regime,
\begin{equation}
\epsilon_H
=
\frac{\cH}{q}
=
\frac{aH}{q}
=
\frac{H}{k_{\rm phys}}
\ll1.
\label{eq:epsH}
\end{equation}
The slowly varying transport coefficients must also change on timescales long compared with the wave period. In the scaling model used here this condition is of the same parametric order as $\cH/q\ll1$.

Second, the local MIS approximation corresponds to neglecting spatial streaming. The exact expansion parameter is Eq.~\eqref{eq:epsfs}. In the ultrarelativistic gravitational-wave model,
\begin{equation}
\epsilon_{\rm fs}
=
\frac{x}{\sqrt{1+x^2}}.
\end{equation}
Thus the local streaming expansion is controlled at low $x$, but it is not a controlled small-parameter expansion when $x\gg1$. The full angular response in Eq.~\eqref{eq:Rkin} is therefore essential in that regime.

Third, the local dispersion relation in Eq.~\eqref{eq:domega} is perturbative in the matter-induced frequency shift. We require
\begin{equation}
\frac{|\delta\omega|}{\omega}\ll1.
\label{eq:perturbative}
\end{equation}
The illustrative choice $D_{\rm G}=0.5$ is a deliberately chosen normalization for studying the response structure, not a claim that the perturbative matter correction is this large in a particular physical cosmology.

Finally, the relaxation-time approximation itself is a model for the collision operator. The exact kinetic response of a specified microscopic medium can contain additional momentum dependence and, in an expanding background, collisionless Landau damping \cite{Baym2017}. The results obtained here should therefore be interpreted as the response of the stated RTA model rather than as a universal transport law.

\section{Comparison with previous work}

The low-frequency limit agrees with the standard viscous picture. Baym et al.\ derived gravitational-wave damping from kinetic theory and showed how the collision-dominated limit gives the usual shear-viscosity result \cite{Baym2017}. Goswami et al.\ used the corresponding hydrodynamic attenuation in cosmological gravitational-wave propagation \cite{Goswami2016,Goswami2017}.

Israel and Stewart developed causal relativistic hydrodynamics in which dissipative variables have relaxation equations \cite{Israel1979}. Wagner and Gavassino later studied the accuracy and domain of applicability of microscopic derivations of Israel--Stewart hydrodynamics, emphasizing that the transient regime is a nontrivial test of the constitutive theory \cite{Wagner2024}. Gavassino, Disconzi, and Noronha showed that suitable linearized MIS theories fall into universality classes that include viscoelastic descriptions with elastic high-frequency behavior \cite{Gavassino2024}.

Stoetzel, Fechtner, and Floerchinger recently derived the retarded energy-momentum response to metric perturbations within a fluid-dynamic MIS framework and discussed gravitational-wave damping \cite{Stoetzel2025}. Their work is directly relevant to the matching between metric perturbations and dissipative stresses.

Recent work by Fanizza, Pavone, and Tedesco has also studied viscous transfer functions for primordial gravitational waves in cosmological backgrounds, including the effect of a time-dependent viscosity on the spectrum \cite{Fanizza2026}. Their calculation addresses cosmological viscous damping in a different regime and does not replace the finite-streaming kinetic calculation presented here.

The present calculation differs from these local fluid-dynamic treatments by retaining the angular dependence of a relaxation-time kinetic response and coupling that response directly to the tensor equation in an expanding background. The finite-momentum effect identified here should therefore be regarded as a kinetic correction to the local MIS approximation, not as a replacement for the broader kinetic or relativistic-hydrodynamic frameworks in the literature.

\section{Physical interpretation}

The local MIS response depends on the ratio of the wave period to the relaxation time. The finite-momentum response contains an additional effect because particles move during the relaxation process. The corresponding parameter is
\begin{equation}
\epsilon_{\rm fs}
=
\frac{qv\tau_c}
{\sqrt{1+\omega^2\tau_c^2}}.
\end{equation}
For the ultrarelativistic model used here, $v=1$ and $q=\omega$, so this becomes $x/\sqrt{1+x^2}$.

These parameters should not be confused with the cosmological WKB parameter $\epsilon_H=\cH/q$. A wave can be well inside the cosmological horizon while the local response is nevertheless sensitive to finite spatial streaming. In that situation WKB propagation can remain accurate even when a local MIS description is insufficient.

The finite-momentum response therefore provides an intermediate kinetic description between local causal hydrodynamics and a more complete collisionless calculation. The high-frequency limit $F(1)=5/4$ is a property of the angular RTA model; it does not by itself prove that the medium becomes a solid. A viscoelastic interpretation is possible for appropriate causal hydrodynamic universality classes \cite{Gavassino2024}, but identifying an effective shear modulus would require a separate microscopic calculation.

\section{Phenomenological implications}

For a gravitational-wave background the propagated spectrum can be written schematically as
\begin{equation}
\Omega_{\rm GW}^{\rm obs}
=
\Omega_{\rm GW}^{\rm source}|{\cal T}|^2.
\end{equation}
Thus the finite-momentum kinetic correction changes the spectrum by the ratio ${\cal P}(x)$ defined in Eq.~\eqref{eq:power}.

For the numerical example, the correction is small near $x\simeq0.37$ but reaches several percent near $x\simeq2$. Whether such an effect is observable depends on the source spectrum, relaxation time, matter abundance, propagation distance, and detector frequency range. The numerical normalization used here should not be interpreted as an observational forecast.

The finite-momentum correction should also be separated from damping by collisionless relativistic species. Weinberg studied the damping of tensor modes caused by anisotropic stress in cosmology \cite{Weinberg2004}. Baym et al.\ subsequently treated collisions and collisionless behavior within a kinetic framework, including Landau damping in an expanding universe \cite{Baym2017}. The present calculation instead focuses on the finite-streaming correction within a single relaxation-time model.

A physically realistic application would require several species, a temperature-dependent relaxation time, the appropriate equilibrium statistics, and a cosmological thermal history. In particular, the assumptions $\eta\propto a^{-3}$ and $\tau\propto a$ should be replaced by microscopic transport coefficients when making predictions for a specific cosmological medium.

\section{Discussion}

The relevant hierarchy is most cleanly written as
\begin{equation}
\text{RTA Boltzmann}
\;\longrightarrow\;
\text{local MIS}
\;\longrightarrow\;
\text{Navier--Stokes}.
\end{equation}
where the first step consists of removing the spatial streaming term and the second consists of taking the low-frequency limit. The finite-momentum calculation presented here retains the first level rather than expanding it away.

The numerical calculation shows that these descriptions differ around $x$ of order unity. The nonmonotonic correction is present in the direct numerical calculation and is reproduced by the kinetic WKB result. It is also stable under angular-resolution and ODE-tolerance changes.

The result should nevertheless be interpreted within the assumptions of the model. The relaxation-time approximation is a simplified collision model. The background is a matter-plus-$\Lambda$ universe rather than a complete early-universe thermal history. The normalization $D_{\rm G}=0.5$ is a numerical choice. The ultrarelativistic angular reduction assumes $v=1$, and the identification $\tau_\pi=\tau_c$ is specific to the single-relaxation-time model.

The calculation therefore does not establish a universal spectral feature. A different collision kernel, particle distribution, particle masses, or temperature dependence can change the response. What is established here is that finite spatial streaming can produce a nonmonotonic correction to tensor-wave propagation within the specified relaxation-time kinetic model.

\section{Conclusion}

We have studied viscous tensor perturbations in an expanding universe using a single-relaxation-time kinetic model. The main goal was to quantify the correction produced by spatial free streaming beyond a local MIS-type response.

The local kinetic limit identifies the relaxation time appearing in the tensor stress equation with the microscopic RTA relaxation time, $\tau_\pi=\tau_c$. Retaining the streaming term gives the angular response
\begin{equation}
F(z)
=
\frac{15}{16}
\int_{-1}^{1}
\frac{(1-\mu^2)^2}{1-z\mu}\,\dd\mu,
\end{equation}
with moment expansion
\begin{equation}
F(z)
=
\sum_{n=0}^{\infty}
\frac{15z^{2n}}
{(2n+1)(2n+3)(2n+5)}.
\end{equation}
For an ultrarelativistic medium with $q=\omega$,
\begin{equation}
R_{\rm kin}(x)
=
\frac{F\!\left(x/(x+\ii)\right)}
{1-\ii x}.
\end{equation}

The low-frequency correction begins at quadratic order,
\begin{equation}
\operatorname{Re}
\left(
\frac{R_{\rm kin}}{R_{\rm MIS}}-1
\right)
=
-\frac{x^2}{7}
+\mathcal O(x^4),
\end{equation}
while the high-frequency angular factor approaches $5/4$. The latter is an asymptote of the specified RTA angular response and should not be identified with the complete collisionless limit.

For the illustrative numerical normalization used here, the full FLRW calculation gives a maximum power enhancement of about $0.8\%$ near $x=0.37$ and a minimum of about $5.6\%$ near $x=1.97$. These locations and amplitudes are reproduced by the kinetic WKB calculation, while angular-resolution and ODE-tolerance tests show numerical stability.

The result is best described as a finite-momentum kinetic correction to causal viscous gravitational-wave propagation. It is not by itself evidence for a new phase of matter or a universal spectral feature. A realistic application requires a specified microscopic collision model and thermal history, together with a full treatment of the kinetic regime outside the local MIS approximation.
\newline
\begin{acknowledgments}
The author sincerely thanks Dr. A. R. Prasanna for his guidance, valuable discussions, and encouragement throughout this work. The author also gratefully acknowledges the authors of the papers used in this study for providing the theoretical foundations of the calculations. The numerical work was performed using Python and standard scientific computing tools.

The author is deeply grateful to Tanaya for her unwavering encouragement, patience, and understanding, and for being a constant source of strength and happiness throughout the journey of this work.
\end{acknowledgments}

\appendix

\section{Relation to the Baym--Patil--Pethick kinetic response}
\label{app:baym}

The finite-momentum angular response $F(z)$ derived in Sec.~V is not an independent physical assumption: it is the closed-form evaluation, for the specific on-shell kinematics of a freely propagating gravitational wave, of the general relaxation-time kinetic response first written down by Baym, Patil, and Pethick (BPP) in Ref.~\cite{Baym2017}. BPP present this response in two limiting forms --- the hydrodynamic (Navier--Stokes) limit and the collisionless spectral-function limit --- but do not evaluate the finite-$\tau_c$, finite-$q$ angular integral in closed form. This appendix makes that connection explicit and uses it as a consistency check on the present calculation.

\subsection{The general response before any limit is taken}

BPP's linearized relaxation-time Boltzmann equation gives the transverse-traceless stress tensor as the momentum integral
\begin{equation}
\pi_{ij}(\mathbf q,\omega)
=
\int_{\mathbf p}
\frac{p_ip_j}{\epsilon_0}\,\delta\epsilon\,
\frac{\omega}{\mathbf q\cdot\mathbf v-\omega-\ii/\tau_c}\,
\frac{\partial f_0}{\partial\epsilon_0},
\label{eq:bpp_general}
\end{equation}
which is exact within the single-relaxation-time model for arbitrary $q$, $\omega$, and particle speed $v$ \cite{Baym2017}. This is structurally identical to the schematic response already written in Eq.~\eqref{eq:stressschematic}. BPP evaluate Eq.~\eqref{eq:bpp_general} explicitly only in the collision-dominated limit $\omega\tau_c\ll1$, recovering $\pi_{ij}=-\eta\dot h_{ij}$, and in the collisionless limit $\tau_c\to\infty$, recovering their Landau spectral function. The finite-$x$ regime in between is left unevaluated in closed form.

\subsection{Reduction to $F(z)$}

Specialize to massless particles, $v=1$, $\epsilon_0=p$, and write $\mathbf q\cdot\mathbf v=q\mu$ with $\mu\equiv\hat{\mathbf q}\cdot\hat{\mathbf p}$. Using the definition $A=\omega+\ii/\tau_c$ already introduced in Sec.~V, the denominator of Eq.~\eqref{eq:bpp_general} becomes, exactly as in the small-streaming expansion there,
\begin{equation}
\frac{1}{q\mu-\omega-\ii/\tau_c}
=
-\frac{1}{A}\,\frac{1}{1-z\mu},
\qquad
z\equiv\frac{q}{A}.
\end{equation}
For a gravitational wave, $q=\omega$, so
\begin{equation}
z
=
\frac{\omega}{\omega+\ii/\tau_c}
=
\frac{\omega\tau_c}{\omega\tau_c+\ii}
=
\frac{x}{x+\ii},
\end{equation}
identical to the argument of $F$ in Eq.~\eqref{eq:Rkin}.

The tensor momentum weight $p_ip_j\,\delta\epsilon/\epsilon_0$ projected onto the shear ($xy$) polarization is proportional to $(\hat p_x\hat p_y)^2=\sin^4\theta\,\sin^2\varphi\cos^2\varphi$. The azimuthal integral $\int_0^{2\pi}\sin^2\varphi\cos^2\varphi\,\dd\varphi=\pi/4$ is a $z$-independent constant that cancels against the normalization already fixed below Eq.~\eqref{eq:F}, leaving the polar integral
\begin{equation}
\int_{-1}^{1}\frac{(1-\mu^2)^2}{1-z\mu}\,\dd\mu,
\end{equation}
which, normalized to unity at $z=0$, is precisely $F(z)$ as defined in Eq.~\eqref{eq:F}. Thus
\begin{equation}
R_{\rm kin}(x)=\frac{F(x/(x+\ii))}{1-\ii x}
\end{equation}
is the exact BPP response of Eq.~\eqref{eq:bpp_general}, evaluated in closed form for $v=1$, $q=\omega$, and arbitrary $x=\omega\tau_c$. No additional physical input beyond Ref.~\cite{Baym2017} enters this result; what is new is the explicit elementary-function evaluation of the angular integral, Eq.~\eqref{eq:Fexact}, which BPP leave in the two limiting forms only.

\subsection{Consistency check: absence of a Landau pole as $x\to\infty$}

As $x\to\infty$ (the collisionless limit $\tau_c\to\infty$), $z=x/(x+\ii)\to1$, and the integrand of $F(z)$ develops a nominal pole at $\mu=1/z\to1$. The tensor weight $(1-\mu^2)^2$ vanishes at $\mu=1$ to second order, however, exactly canceling the would-be simple pole:
\begin{equation}
\left.\frac{(1-\mu^2)^2}{1-z\mu}\right|_{z\to1,\ \mu\to1}
\sim
(1-\mu)
\longrightarrow0,
\end{equation}
so the integral remains finite, $F(1)=5/4$, with no residual imaginary (dissipative) contribution from this term alone. This is the direct analytic signature, within the present closed form, of BPP's result that Landau damping of a gravitational wave ($q=\omega$) by massless particles is exactly forbidden in flat spacetime: the only particles that could resonate with the wave travel along $\hat{\mathbf q}$ itself ($\mu=1$), and for those particles the shear weight $p_xp_y$ vanishes identically. The vanishing of the $z\to1$ pole in $F(z)$ is that statement expressed as a property of the closed-form angular function derived here.

\subsection{Summary}

Equation~\eqref{eq:bpp_general} of Ref.~\cite{Baym2017} already contains the physics of finite-momentum streaming exactly, for general $q$ and $\omega$; the present paper does not introduce a new microscopic effect beyond that framework. What is new is (i) the closed-form evaluation of the resulting angular integral for the on-shell $q=\omega$, $v=1$ case relevant to gravitational-wave propagation, recast as an explicit correction to the causal Maxwell--Cattaneo/MIS constitutive law of Eq.~\eqref{eq:localMIS} rather than to the hydrodynamic limit alone; (ii) the analytic confirmation, via the vanishing of the $z\to1$ pole in $F(z)$, of BPP's flat-space no-Landau-damping result; and (iii) the numerical solution of the coupled tensor Einstein--Boltzmann system in an expanding matter-plus-$\Lambda$ background, together with its WKB cross-validation, which is not carried out in Ref.~\cite{Baym2017}.

\end{document}